# Realizing Quantum Algorithms on Real Quantum Computing Devices

Carmen G. Almudever*, Lingling Lao*, Robert Wille†, Gian Giacomo Guerreschi‡

* Quantum & Computer Engineering Department and QuTech, Delft University of Technology
† Institute for Integrated Circuits, Johannes Kepler University Linz
‡ Intel Labs

***Abstract*—Quantum computing is currently moving from an academic idea to a practical reality. Quantum computing in the cloud is already available and allows users from all over the world to develop and execute real quantum algorithms. However, companies which are heavily investing in this new technology such as Google, IBM, Rigetti, Intel, IonQ, and Xanadu follow diverse technological approaches. This led to a situation where we have substantially different quantum computing devices available thus far. They mostly differ in the number and kind of qubits and the connectivity between them. Because of that, various methods for realizing the intended quantum functionality on a given quantum computing device are available. This paper provides an introduction and overview into this domain and describes corresponding methods, also referred to as compilers, mappers, synthesizers, transpilers, or routers.**

## I. Introduction

Quantum computing has been a very active and promising area of research and, especially in the last years, of technology development. Since the physicist Richard Feynman proposed the idea of building a quantum computer to simulate quantum systems in the early 80's [1], several quantum algorithms and quantum error correction techniques have been developed [2], [3]. By exploiting quantum phenomena such as superposition and entanglement, quantum computers promise to solve hard problems that are intractable for even the most powerful conventional supercomputers. In addition, remarkable progress has been made in quantum hardware based on different technologies such as superconducting circuits, trapped ions, silicon quantum dots, and topological qubits [4]–[7]. A recent breakthrough in quantum computing has been the experimental demonstration of quantum supremacy[1] using a superconducting quantum processor consisting of 53 qubits [8].

Current and near-term quantum computing devices are often referred to as *Noisy Intermediate-Scale Quantum* (NISQ) devices [9], to highlight their limited size and imperfect behaviour due to noise. However, while quantum technologies need to improve coherence times and gate fidelities to achieve overall lower error rates, quantum computing in the cloud is already a reality offering small quantum computing devices that are capable of handling basic quantum algorithms. Companies such as Google, IBM, Rigetti, and Intel, have already announced 72-qubit, 50-qubit, 128-qubit, and 49-qubit superconducting devices, respectively.

In these quantum processors, qubits are arranged in a 2D topology with limited connectivity between them and in which only nearest-neighbor (NN) interactions are allowed. This is one of the main constraints of today's quantum devices and frequently requires the quantum information stored in the qubits to be moved to other adjacent qubits – typically by means of SWAP operations. Quantum algorithms, which are described in terms of quantum circuits, neglect the specific qubit connectivity and, therefore, cannot be directly executed on the quantum computing device but need to be realized with respect to this and others constraints. The procedure of adapting a circuit to satisfy the quantum processor restrictions is known as the compiling, mapping, synthesis, transpiling, or routing problem.

The mapping process often causes an increase of the number of quantum operations as well as the depth (number of time-steps) of the quantum circuit. The success rate of the algorithm is consequently reduced since quantum operations are error prone and qubits easily degrade their state over the time due to the interaction with the environment. To minimize the negative impact of the mapping, it is required to develop efficient methods that minimize the resulting overhead – especially for NISQ devices in which the lack of active protection against errors will make long computations unreliable.

This paper will provide an introduction and overview on the mapping problem and, by this, on how to realize quantum algorithms that are represented in terms of quantum circuits on real quantum devices. To this end, we will first review the basics of quantum computing and, afterwards, will present two mapping approaches for realizing quantum algorithms on two different superconducting transmon devices. The first targets an IBM processor, the *IBM QX4* [10], that consists of five qubits. The second is meant to execute quantum circuits on the *Surface-17* chip [11], [12] composed of seventeen quits.[2] A more detailed discussion on the different kind of quantum devices and the internal representations required by the mappers will also be provided. Finally, we will pose some open questions that the quantum compilation community should consider.

[1]A quantum computer is capable of solving a computational task that would require an unreasonable amount of time on any classical supercomputer.

[2]Note that larger quantum architectures are available from both vendors, but to keep the following descriptions and examples simple, we use these ones.

## II. Basics on Quantum Computing

In contrast to classical circuits and systems, computations in the quantum realm rely on so-called *quantum bits* or *qubits*. Qubits can assume the well-known basis states $|0\rangle$ and $|1\rangle$ (here written using Dirac notation), but can also be put into *superposition* of both. More precisely, the state of a qubit (in other words, a *quantum state*) can be described by $|\psi\rangle = \alpha_0 |0\rangle + \alpha_1 |1\rangle$, where $\alpha_0$ and $\alpha_1$ are complex numbers called *amplitudes* and $|\alpha_0|^2 + |\alpha_1|^2$ has to be equal to 1. Measuring a single qubit in the computational basis will result in a binary value, 0 or 1, collapsing the qubit to either of the two basis states $|0\rangle$ and $|1\rangle$ with probability $|\alpha_0|^2$ and $|\alpha_1|^2$, respectively. The state of $n$ such qubits is described by the tensor product of the individual states – eventually leading to a state described by $2^n$ amplitudes $\alpha_{0\ldots0}, \alpha_{0\ldots1}, \ldots, \alpha_{1\ldots1}$, which is usually provided in terms of a state vector.

A quantum state can be changed by applying *quantum gates* on it. Each quantum gate can be described by a *unitary matrix* and may act on one or more qubits, although usually only one-qubit and two-qubit gates are naturally supported by most quantum computing devices. Common gates performed on single qubits are the *Hadamard* gate $H$ to set a qubit into superposition, the *Pauli* gates $X$, $Y$, and $Z$ which rotate the qubit state on the respective axis (assuming a Bloch-sphere description of the qubit state [13]), as well as the *phase shift* gate $T$. They are described by

$$H = \frac{1}{\sqrt{2}}\begin{bmatrix}1 & 1\\ 1 & -1\end{bmatrix}, \quad X = \begin{bmatrix}0 & 1\\ 1 & 0\end{bmatrix},$$

$$\mathrm{Y} = \begin{bmatrix}0 & -i\\ i & 0\end{bmatrix}, \quad \mathrm{Z} = \begin{bmatrix}1 & 0\\ 0 & -1\end{bmatrix}, \text{ and } \quad \mathrm{T} = \begin{bmatrix}1 & 0\\ 0 & e^{\frac{i\pi}{4}}\end{bmatrix}.$$

Two-qubit operations exist, e.g., in terms of *controlled* versions of single-qubit gates, where one qubit acts as *control qubit* and the other one acts as *target qubit* – eventually, employing, e.g.,

$$\mathrm{CX} = \begin{bmatrix}1 & 0 & 0 & 0\\ 0 & 1 & 0 & 0\\ 0 & 0 & 0 & 1\\ 0 & 0 & 1 & 0\end{bmatrix} \text{ and } \quad \mathrm{CZ} = \begin{bmatrix}1 & 0 & 0 & 0\\ 0 & 1 & 0 & 0\\ 0 & 0 & 1 & 0\\ 0 & 0 & 0 & -1\end{bmatrix}$$

for the *controlled* $X$ gate, also known as *controlled* NOT gate abbreviated with CNOT, since $X$ realizes a NOT operation, and the *controlled* $Z$ gate, respectively. Besides that, the SWAP gate defined by

$$\mathrm{SWAP} = \begin{bmatrix}1 & 0 & 0 & 0\\ 0 & 0 & 1 & 0\\ 0 & 1 & 0 & 1\\ 0 & 0 & 0 & 1\end{bmatrix}$$

exchanges the values of the two involved qubits, which is essential for the mapping methods described in this paper.

To evaluate the effect of a quantum gate on a quantum state, the respective vector (describing the quantum state) simply has to be multiplied with the respective matrix (describing the gate). The gates reviewed above form a universal gate set, i.e., all quantum functions can be realized by them.

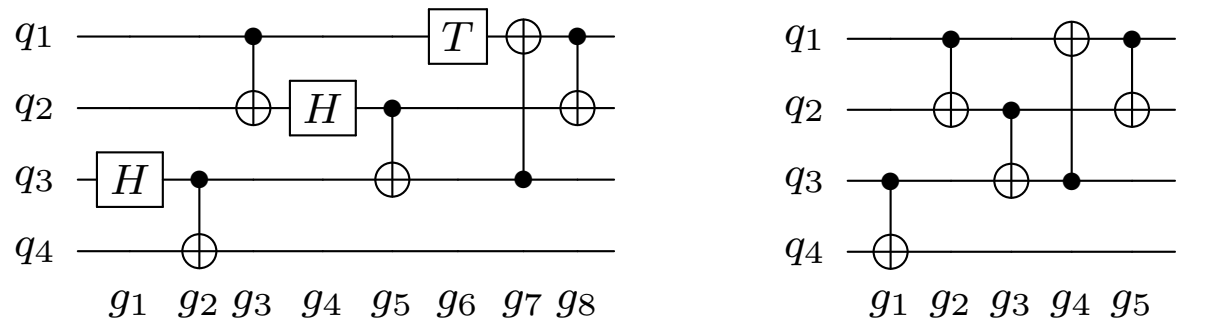

**(a)** W/ all gates **(b)** W/o single-qubit gates

Figure 1: An example of a quantum circuit.

Sequences of quantum operations finally define quantum algorithms which are usually described by high-level quantum languages (e.g. Scaffold [14] or Quipper [15]), quantum assembly languages (e.g. OpenQASM 2.0 developed by IBM [16] or cQASM [17]), or circuit diagrams.

For the purpose of this overview, we are using circuit diagrams such as those in Fig. 1(a) as representation of quantum algorithms in the following. Here, qubits are visualized as circuit lines that are passed through quantum operations, which are denoted by boxes including their respective denominator in case of single-qubit operations and a black dot and a $\oplus$-symbol for control qubit and target qubit, respectively, in case of a CNOT operation. Note that the qubit lines do not refer to an actual hardware connection as in classical logic, but rather define in which order (from left to right) the respective operations are applied.

Physical implementations of quantum computers may rely on different technologies. In this work, we will focus on quantum computers based on superconducting transmon qubits [18] on silicon chips. Here, operations are conducted through microwave pulses transferred into and out of dilution refrigerators, in which the quantum chips are set at an operating temperature of around 15 mK. Communication into, out of, and among the qubits is done through on-chip resonators.

## III. Mapping Quantum Circuits to Quantum Computing Devices

In this section, we review the mapping problem and provide a brief overview of selected previous work on the topic. Based on that, the following sections provide descriptions of mapping methods that have explicitly been developed for existing quantum devices as well as a more unifying look to this mapping problem.

### A. The Mapping Problem

As in classical computers, quantum algorithms described as programs using a high-level language have to be compiled into a series of low-level instructions like assembly code and, ultimately, machine code. As sketched in Figure 2, in a quantum computer these instructions need to be ultimately translated into the pulses that operate on the qubits and perform the desired operation [19].

In this context quantum algorithms can be described as a list of sequential gates, each acting on a few qubits only, and visualized in terms of quantum circuits. Quantum circuits cannot be directly realized on real quantum processors, but need to be adapted to the specificity of each quantum device. In addition to preserving all dependencies between the quantum operations, compilers of quantum circuits must perform three important tasks: 1) express the operations in terms of the gates native to the quantum processor, a task called gate decomposition, 2) initialize and maintain the map specifying which physical qubit (qubit in the quantum device) is associated to each program qubit (qubit in the circuit description, sometimes called logical qubit in the literature), a task called placement of the qubits, and 3) schedule the two-qubit gates compatibly with the physical connectivity, often by introducing additional routing operations.

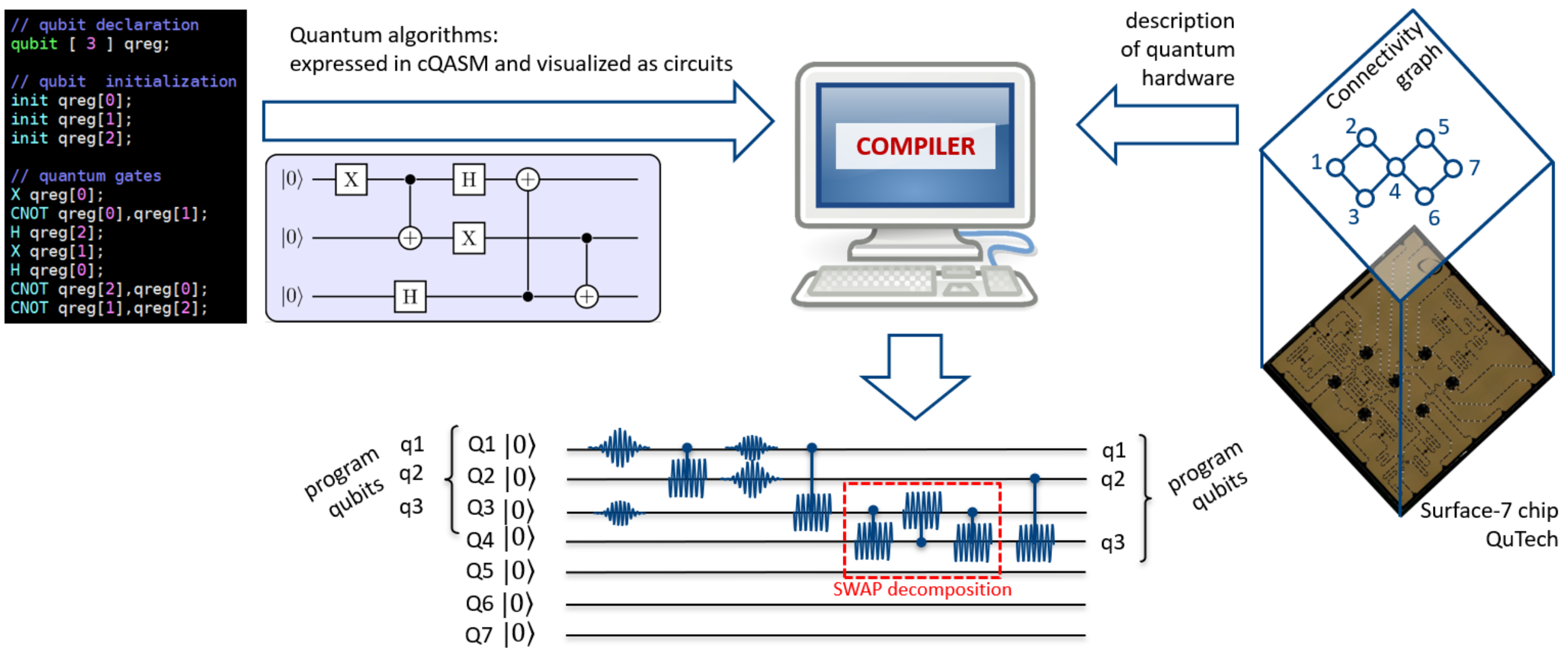

Figure 2: Sketch of the mapping process for quantum algorithms. The compiler depicted in the center receives two kinds of inputs: from the left it receives the quantum algorithm in terms of a sequential list of quantum gates to be executed (expressed in cQASM [17]) and from the right a description of the machine, possibly including the control electronics in addition to the quantum hardware. Its output is a series of scheduled operations that can be executed by the machine and is depicted at the bottom in terms of the control signals that implement it. The initial placement of the program qubits $\{q_1, q_2, q_3\}$ may differ from the final placement. For simplicity we have assumed that the CNOT and H gates are available in the machine's gate set instead of the native gates of Surface-7.

In this work we do not elaborate on the gate decomposition, apart from observing that most current quantum devices provide a native gate set that is equivalent and often larger than the universal gate set described in the previous section.[3] The two remaining tasks are performed by the circuit mapper within the compiler. Notice that the task of initializing the qubit placement is expected to play an important role in near term devices, but will probably have a relatively limited impact when algorithms grow in depth. For this reason, the main focus of the following sections will be on the problem of minimizing the routing overhead, arguably the most impactful mapping task especially when excluding quantum error correction.

The problem is simply stated: one needs to schedule a two-qubit gate but the corresponding program qubits are currently placed on non-connected physical qubits. The placement must therefore be modified with the goal of moving the involved qubits to adjacent connected ones. Quantum information cannot be copied and there is essentially one way of transferring it[4], namely by applying SWAP gates that effectively exchange the state of two connected qubits.

The functionality of the circuit mapper, which is usually embedded in the compiler, is sketched in Fig. 2. It requires two separate inputs, one related to the abstract algorithm to implement and the other associated with the quantum processor chosen for its execution. The former is usually provided in terms of high-level code [14], [15] or Quantum Assembly Language (QASM) instructions [16], [17], an explicit list of low-level operations corresponding to single- and two-qubit gates, and can be visualized in the form of quantum circuits. The latter corresponds to a description of the hardware, from the qubit topology and connectivity to the electronics that generate and distribute the control signals. The compiler is in charge of decomposing the operations in terms of the gates native to the processor and then of the mapping process. The mapping process is comprised of the initial placement of qubits, qubit routing, and operation scheduling.

## B. Prior Work

Several solutions have already been proposed for solving the mapping problem; that is, to make quantum circuits executable on the targeted quantum device by transforming and adapting them to the constraints of the quantum processor [24]–[54]. Most of the works focus on NISQ devices such as the IBM [10] or Rigetti [55] chips as they are accessible through the cloud. The proposed mapping solutions differ and therefore can be classified according to the following characteristics:

- **Quantum hardware constraints:** one of the main restrictions of current quantum devices is the limited connectivity between the qubits. Different quantum processors, even within the same family, can have different topologies such as a linear array (1D), a 2D array with only nearest-neighbour interactions [29], [30], [32], [38], [41], or more arbitrary shapes [36], [42], [43], [48], [52]. Although most of the works on mapping focus on the qubit connectivity constraint, there are other restrictions that originate from the classical control part and that reduce the parallelizability of quantum gates [35], [39]. This kind of constraint become more and more relevant when scaling-up quantum systems as resources need to be shared among the qubits.
- **Solution approach and methodology:** exact approaches [30], [43], [49] are feasible when considering relatively small number of qubits and gates, giving minimal or close-to-minimal solutions. However, they are not scalable. Approximate solutions using heuristics can be used for large quantum circuits [25], [34], [52]. Some used methods are (Mixed) Integer Linear Programming ((M)ILP) solvers [24], [39], [53], Satisfiability Modulo Theory (SMT) solvers [43], [45], [46], heuristic (search) algorithms [24], [28], [31], [33], [40], [44], [54], decision diagrams [27], or even temporal planners and reinforcement learning [37], [51].
- **Cost function:** there are different metrics that can be optimised in the mapping process. The most common cost functions are the number of gates (i.e. minimize the number of added SWAPs) and the circuit depth or latency (i.e. minimize the number of time-steps of the circuit). Recent works started optimising directly for circuit reliability (i.e. minimize the error rate by choosing the most reliable paths) [45]–[47], [50].
- **Solution features:** In addition to the just mentioned characteristics, there are other important features that can lead to better solutions. Some examples are the look-back strategy in the routing that refers to taking into account the previous already scheduled operations when selecting the routing path [39] or the look-ahead feature [54] [40], [52] that considers not only the current two-qubit gates that need to be routed and scheduled but also some of the future ones with some weights. Besides that, also pre-processing steps dedicated to particular quantum functionality have shown to be extremely beneficial [26].

[3] Approaches for decomposition have, e.g., been introduced in [20]–[23].

[4] Another approach is based on teleportation, corresponding to long-distance transfer of the qubit state. It requires the creation of multi-qubit entangled states that are preliminarily distributed across the qubit register and that can be consumed to transfer a qubit state. Since the distribution of the entangled state requires SWAP gates, the teleportation approach can be seen as a SWAP-based routing with relaxed time constraints.

In the following sections, we will describe some of these mapping methods, which have explicitly been developed for current quantum processors.

## IV. Mapping Quantum Circuits on IBM Q Devices

In 2017, IBM launched the *IBM Quantum Experience* [?]) – a web portal which allows users to write quantum programs and run them on actual quantum computers. To this end, physical realizations of quantum computers have been made publicly available through cloud access. Diagrams of the various IBM Q quantum chips are available in [10].

Those implementations support the elementary single qubit gates $U(\theta, \phi, \lambda) = R_z(\phi)R_y(\theta)R_z(\lambda)$ (i.e. an Euler decomposition) that is composed by two rotations around the $z$-axis and one rotation around the $y$-axis, as well as the CNOT operation. By adjusting the parameters $\theta$, $\phi$, and $\lambda$, single-qubit gates of other gate libraries like the $H$ or the $T$ gate can be realized (among others like rotations). All other gates and particularly all gates acting over more than two qubits such as the Toffoli operation or the Fredkin operation, have to be decomposed into those native gates. To this end several methods have been proposed in the literature (see, e.g., [20]–[23]).

Besides that, however, also so-called coupling or connectivity restrictions have to be satisfied. This affects the two-qubit CNOT gates which cannot arbitrarily be placed because, in IBM's implementations, they are

- allowed to interact between dedicated pairs of qubits only and,
- within these interactions, have to follow a firmly defined scheme of which qubit may work as target and which qubit may work as control.

More precisely, for each IBM QX quantum architecture, a so-called *coupling graph* is provided which defines the allowed interactions. Nodes of the graph indicate physical qubits (denoted by $Q_i$), while directed edges define the possible CNOT applications, i.e. an edge pointing from physical qubit $Q_i$ to qubit $Q_j$ defines that a CNOT with control qubit $Q_i$ and target qubit $Q_j$ can be applied. All other interactions are prohibited.

Fig. 3(a) shows the coupling graph and, by this, the allowed qubit interactions for IBM's *IBM QX4* device. Because of that, the circuit considered before in Section II and shown in Fig. 1 cannot be directly executed on this device if the program qubits $q_1, q_2, q_3, q_4$ are placed (mapped) to the physical qubits $Q_1, Q_2, Q_3, Q_4$ of the architecture. This is because, for instance, the first CNOT gate works with qubit $q_3$ as control and qubit $q_4$ as target which is not allowed according to the coupling graph.

A straight-forward approach to solve this problem is to reposition the qubits by SWAP gates, which afterwards have to be decomposed into native gates. By this, qubits can be "moved" to positions in which their interactions are allowed. For example, the circuit shown in Fig. 3(b) uses additional SWAP gates so that it now realizes the original circuit from Fig. 1 but, at the same time, is compliant to all constraints imposed by the coupling graph. Note that, SWAP gates are drawn by two $\times$-symbols. Furthermore, note that, for sake of clarity, we removed all single-qubit gates, since they naively satisfy the coupling constraints. Hence, to see the difference before/after mapping, compare Fig. 1(b) and Fig. 3(b).

However, SWAP gates obviously increase the gate count and the circuit depth and, by this, the costs of the circuit. These further gates increase the chance of errors during a quantum computation. Hence, the main objective of every designer is to keep this overhead as small as possible – an $\mathcal{NP}$-hard problem [56]. In fact, in order to map a quantum

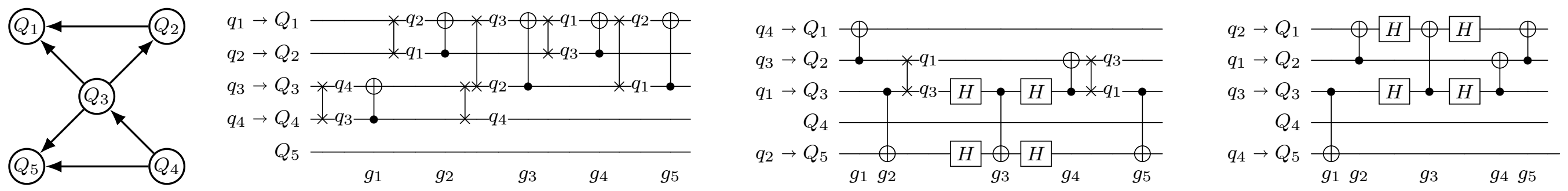

**(a)** Coupling graph **(b)** Naive solution **(c)** Heuristic solution **(d)** Exact solution

Figure 3: Mapping quantum circuits on the *IBM QX4* device.

circuit composed of $n$ program qubits as well $|G|$ CNOT gates to an IBM device with $m$ physical qubits, a total of $2^{n \cdot m \cdot |G|}$ possible combinations have to be checked [57].

Accordingly, IBM itself but also researchers started to investigate more efficient solutions to tackle this problem. This resulted in exact approaches such as introduced in [57], [58] as well as heuristic solutions such as introduced in [26], [40], [49], [54], [58], [59]. While the former can guarantee minimal or, at least, close-to-minimal solutions, they are often not that scalable even though the use of reasoning engines such as SAT solvers yield impressive improvements [57]. Nevertheless, these solutions are still important, e.g., to determine minimal building blocks and to evaluate the quality of heuristic approaches. For actual use cases, however, the heuristic approaches are still the best solution – even if they cannot guarantee a minimal overhead and, in fact, are often far away from the optimum.

As examples, consider again the circuit shown in Fig. 1(b) which shall be realized on an IBM device with a coupling graph as shown in Fig. 3(a). While the naive approach discussed before by means of Fig. 3(b) yields a significant overhead, a heuristic solution shown in Fig. 3(c) (determined using [54]) is significantly cheaper. Here, also $H$ gates are employed to flip the direction of the control and target qubits. Still, even this solution can be further improved as the result of an exact approach shown in Fig. 3(c) (determined using [57]) confirms.

## V. Mapping Quantum Circuits on the Surface-17 Device

Most of the solutions proposed for the mapping problem focus on quantum processors that are available in the cloud, that is, IBM and Rigetti quantum devices. A more scalable quantum processor with a surface code architecture was presented in [11], [12], called *Surface-17*. This quantum chip has been built with the goal of demonstrating fault-tolerant (FT) computation in a large-scale quantum system based on surface code [60], one of the most promising quantum error correction (QEC) codes. However, it can also be considered as a NISQ device and therefore be used for running quantum algorithms that require up to 17 qubits.

The Surface-17 chip is based on superconducting transmon qubits that are operated at very low temperatures ($\sim$ 20 mK). In this implementation, in principle it is possible to perform any kind of single-qubit gates. However, usually gates are limited to a finite set due to the limitation on the amount of gates that can be predefined. In this case, available single-qubit gates are X and Y rotations as they are easier to implement. In addition, the native two-qubit gate is the conditional-phase gate, also called CZ gate (see section II). Therefore, any operation in the quantum algorithm needs to be decomposed to the mentioned native gates before being executed on the Surface-17 processor [39].

As in the IBM chip, the Surface-17 also has some coupling or connectivity restrictions. Its topology is shown in Fig. 4 and corresponds to a 2D array of qubits. Circles represent the physical qubits and the edges the connections (resonators in the real chip) and therefore possible interactions between them. For instance, qubits 1 and 5 can interact, that is, perform a CZ gate, but realising a two-qubit gate between qubits 1 and 7 is not possible. In other words, two-qubit gates can only be performed between nearest neighbouring qubits. Note that in this case, there is no restriction on which qubit can act as a control or as a target. As mentioned in the previous section, qubits can be moved to adjacent positions by using SWAP operations that in Surface-17 chip need to be further decomposed into CZ and Y rotations (see Fig. 6).

Another important limitation in current quantum devices that has not been considered so far in previous mapping works, is the so-called classical control constraint. Superconducting qubits are operated by applying specific microwave pulses [59]. These signals are generated by classical electronics such as Arbitrary Waveform Generators (AWGs) located at room temperature. Qubits could be operated independently by having a dedicated control device for each of them. This would allow, for instance, to perform in parallel any possible combination of single-qubit gates as long as the dependency between the operations was respected. However, this dedicated control approach is not an scalable, feasible and affordable (in terms of cost), specially for building large-scale quantum systems. Therefore, control instruments need to be shared among different qubits. This restriction may severely affect the scheduling of quantum operations as it will limit the possible parallelism leading to larger circuit depths. As previously mentioned, the larger the circuit depth, the lower the algorithm's reliability as the computation time is limited by the coherence time of the qubits.

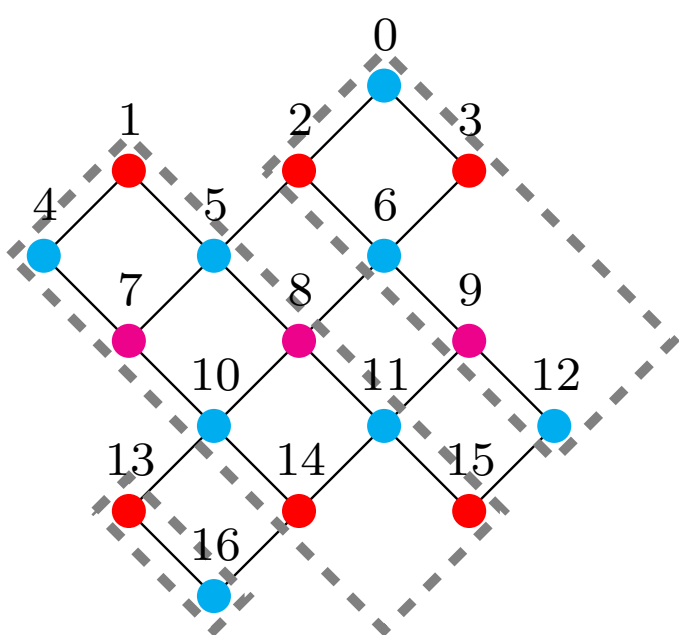

Figure 4: Schematic of the realization of the SC-17 processor.

In Surface-17 chip, single-qubit gates correspond to microwave pulses that are applied at a frequency resonant to the energy of the qubits. The chip's qubits have one of three frequencies, denoted by $f_1$, $f_2$, and $f_3$ (with $f_1 > f_2 > f_3$) and indicated in Fig. 4 by colors red, blue and pink respectively. Assuming a single microwave generator to operate to the same frequency qubits has the following consequence. The same single-qubit gate (e.g. X gate) can be performed in all or some of the same frequency qubits being red or blue or pink. However, different single-qubit gates cannot be applied to the same frequency qubits at the same time because that will require to generate different pulses. That is, an X gate can be applied on all red qubits simultaneously but one cannot perform an X gate on qubit 1 and a Y gate on qubit 2 at the same time. In addition, in this quantum chip several qubits are measured through the same feedline to which they are coupled to. This is illustrated by a dashed grey rectangle in Fig. 4. For instance, qubits 0, 2, 3, 6, 9, and 12 are coupled to the same feedline. Note that measurement takes several cycles. This means that a measurement in all or some of the six qubits can start at the same time but it is not possible to start measuring qubit 2 while still measuring qubit 0. Finally, the implementation of CZ gates also imposes limitations on the parallelism of operations. A CZ gate in Surface-17 is realized by bringing the involved qubits close in frequencies - usually the frequency of the higher frequency qubit is lowered to be close to the qubit with the lower one. In this process, the qubits performing the CZ gate might interact with other neighbouring qubits that share a connection with any of them and are also close in frequencies. In order to avoid such unwanted interactions, qubits have to be detuned to a so-called *parking frequency*. These qubits cannot be involved in any single or two- qubit gate during the time they are in the parking frequency. A more detailed explanation can be found in [39].

In [39] a mapper called *Qmap* for the Surface-17 processor is presented. It is embedded in the OpenQL [61], [62] compiler and it adapts the quantum circuit to the quantum hardware constraints that are described in a configuration file. Note that Qmap can easily target other quantum devices by just changing the parameters in this file. It consists of three blocks: initial placement, qubit routing and operations scheduler. An Integer Linear Programming (ILP) algorithm is used to find an optimal initial placement in which qubits are placed in the chip according to their interactions, whereas an heuristic algorithm is used for the routing task. In this case the cost function (metric to minimize in the routing step) is the circuit latency that refers to the execution time of the algorithm when considering the real gate duration. This means that the routing path that results in the lowest latency overhead and therefore maximises the instruction-level parallelism is selected (looking-back feature).

Considering the circuit shown in Fig. 1(a) and using Qmap to map it into the Surface-17 processor, results in the circuit shown in Fig. 5. After the initial placement of qubits, gates are scheduled and only one SWAP is added to comply to the coupling restrictions. Both single and two-qubit gates are considered and scheduled as Qmap optimizes for circuit latency. Note that the circuit in Fig. 5 still needs to be decomposed to the native gates as shown in Fig. 6 and re-scheduled taking the electronic control constraints into consideration. In this case, the circuit latency will be 26 cycles (20 ns per cycle) that is an ~2x increase compared to the circuit latency before mapping, in which the circuit is decomposed into the native gates and operations are scheduled only considering the dependencies between them.

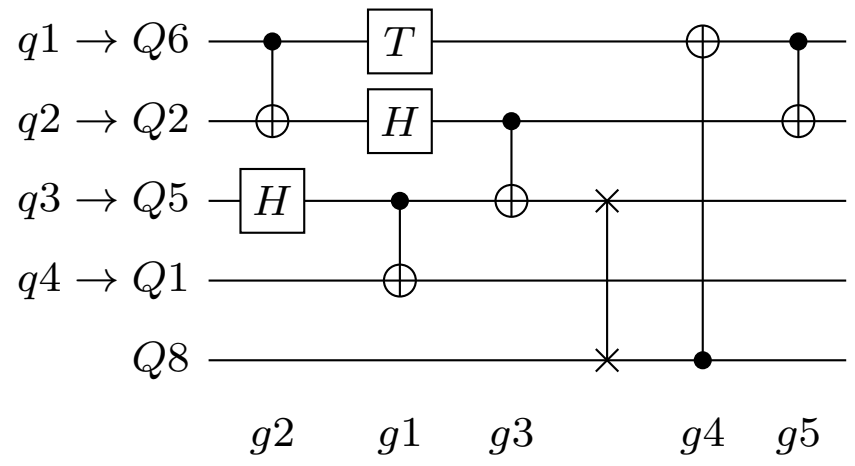

Figure 5: Solution when only considering the connectivity constraint and operations dependency. Gates vertically adjacent can be executed in parallel.

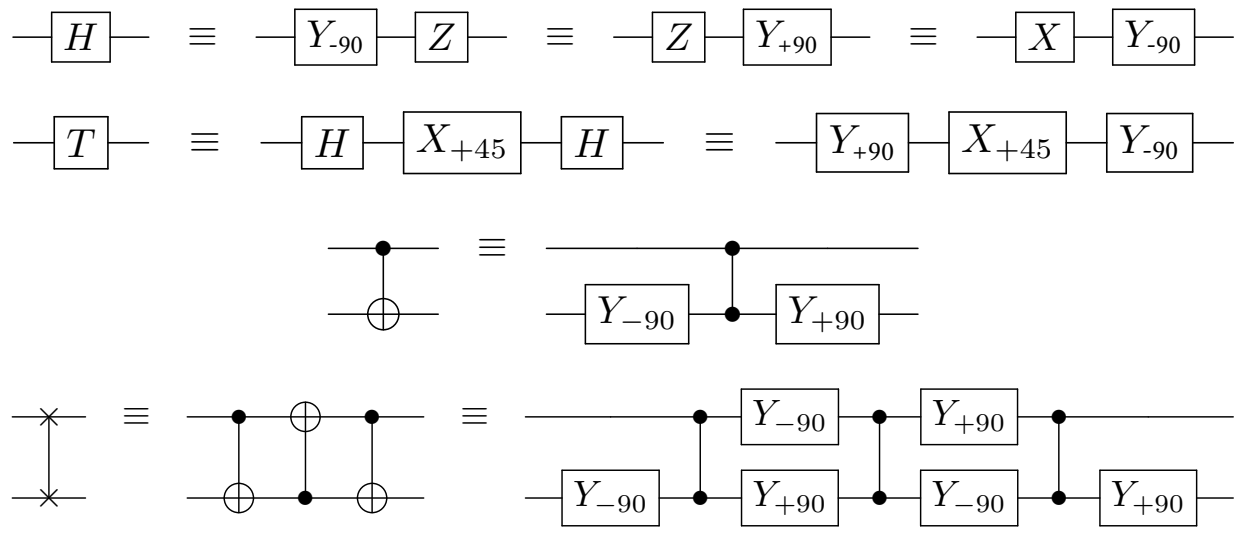

Figure 6: Gate decomposition into native gates supported in the superconducting Surface-17 processor.

## VI. Every Device is (almost) Equal Before the Compiler

In the previous sections we have presented two different approaches to schedule quantum circuits, each developed to satisfy the constraints of a specific quantum device, namely IBM's QX4 and QuTech's Surface-17. In this section we want to provide a more unifying look to the mapper problem and to the description of quantum devices. We start from the latter point.

### A. Device Types

Certain machines allow for extensive pre-compilation of the algorithms that solely excludes the routing operations and parallelization information. In this case the mapper receives QASM code that uses only the one- and two-qubit gates available to the device. The output only adds routing operations. These machines require:

- symmetric two-qubit gates
- homogeneous single-qubit gate set
- the possibility of measuring any qubit in the same basis.

Here, SWAP gates are needed only to overcome the connectivity limitations. The mapper needs to know how to decompose SWAP gates into the available gate set. Often there are multiple ways to do so, for example each decomposition originates a second one obtained by exchanging the role of two qubits involved in the first one. While the transmon architecture of Surface-17 chip exhibits the three properties listed above, see Section V, they are not required for functioning quantum devices. When the properties are not satisfied, the mapper cannot fully separate the gate decomposition and routing tasks.

When the two-qubit gates are asymmetric, decisions concerning the addition of extra gates must be made at the time of routing and scheduling. For example, when CNOTs are used as in the IBM architecture of Section IV, extra Hadamard gates may be required to invert the role of the control and target qubits. This can be known only at the time of routing, i.e. when the qubit placement in the CNOT is known.

When the available native one-qubit gates differ from (physical) qubit to qubit, the scheduling involves two steps. Consider that one needs to schedule gate $U$ acting on the $k$-th program qubit. In this case it is required to 1) compute the sequence of available gates that implements, or approximates, $U$ for the different physical qubits (or at least those at short distance from the physical qubit currently associated to program qubit $k$), and 2) add the cost of the routing. Selecting the better option therefore requires performing multiple gate decompositions and can be done only at scheduling time when the placement is known. To date all architectures provide the same set of one-qubit gates per physical qubit, but this may change due to the pressure of reducing control resources or when the gate fidelity is used as the metric to guide mapping decisions.

Finally, when not all qubits can be directly measured or when the available measurements differ from qubit to qubit, additional gates are required. In the first case to move the quantum state towards measurable qubits, and in the second case to adapt the measurement basis.

### B. Internal Representation Required by Mappers

Despite their differences, all mappers needs an internal representation of key quantities and these can be combined in the concept of the *execution snapshot*. As the name suggests, the execution snapshot is a complete description of the algorithm and its current, usually partial, schedule. It contains:

- the dependency graph of the algorithm with the indication of which gates have already been scheduled
- the initial placement that associate each program qubit to a physical qubit
- the current placement of the qubits
- the partial schedule with the timing information and explicit parallelism
- the settings of the control electronics for the execution.

The data structure specifying the execution snapshot varies from mapper to mapper. Here we provide an intuitive one: the dependency graph is a directed, acyclic graph with nodes representing the quantum gates and edges indicating dependencies (the target node corresponds to the gate that depends on the source node) [34], [44]. Nodes can have one of two colors, differentiating the gates already scheduled from those that need to be scheduled. An additional color may mark the gates that can be scheduled next according to the algorithmic dependencies. Qubit placement is represented by an array of integers of size equal to the number of physical qubits: the $k$-th entry corresponds to the index of the program qubit associated to the $k$-th physical qubit, apart from a special integer indicating that the qubit is "free" in situations where the program requires less qubits than those present in the quantum hardware. Finally the schedule with timing information can be provided as a table by discretizing the time into clock cycles, the greatest common divisor of the gates' duration. This table also includes any additional gate from gate decomposition and routing.

To conclude, the mapper has to take into account the constraints from the control electronics. To this end one needs a way to track, for every clock cycle in which a certain gate can be performed according to the logical dependencies, if that gate can be executed compatibly with all the gates already scheduled. Therefore the mapper needs to be aware of how the set of compatible gates (i.e. those part of the physically available gate set and that do not conflict with gates already scheduled) changes at each clock cycle, and update it dynamically. The conceptually simplest method is to keep an explicit list of the compatible gates for each physical qubit, but this may not be the most efficient implementation. In fact more compact representations are derivable for specific architectures [35], [39].

### C. Unique Hardware Features

In Section IV and V we described two devices based on superconducting circuits. This is by no mean the only approach to scalable quantum devices. In the introduction we mentioned that multiple physical implementation of quantum processors are currently developed, including but not limited to trapped ions, silicon quantum dots, photonics, neutral atoms, and topological systems. The maturity of each technology is at a different point and the challenges to scalability are also different [4]–[7]. Here we are interested to provide a few examples in which particular physical implementations provide unique features. We only present three of them for illustration purposes.

Most architectures are limited to a planar connectivity between qubits, but trapped ions provide all-to-all connectivity, at least inside groups of tens of ions [63]. This is originated by their long-distance Coulomb interaction and mediated by their collective vibrational modes. However this desirable property comes at the price of reduced two-qubit gate parallelism. Finally observe that multi-qubit gates are also available for trapped ions [64] and this may require an enlarged instruction set.

Photonics architectures are uniquely positioned for tasks that combine computation and communication, like at the nodes of quantum repeaters' networks [65], [66]. However they are limited to demolition measurements in which the qubit is "destroyed" when measured since the photon is absorbed by the detector. One can generate a new photon to re-initialize the qubit state.

In silicon quantum dots the role of qubits is played by the spin of electrons confined in electromagnetic potential wells called dots. The simplest scheme is one electron per dot, but alternative configurations are also considered. Two-qubit gates are implemented via the exchange interaction between two electrons in nearby dots [67]. However certain dots can be momentarily empty and electrons can be moved to empty dots in a way that maintains the qubit coherence, the so called shuttling operation [68]. The electron movement can be interpreted either as a change in the device connectivity or as an alternative qubit routing not based on SWAP gates. Specialized mappers are required to take full advantage of these capabilities.

## VII. Conclusions and Discussion

In this paper, we have provided an introduction and overview on the realization of quantum algorithms on real quantum computing devices, and more specifically to the mapping problem. During the compilation process quantum circuits need to be modified to comply with the constraints of the quantum device. This usually results in an increase of the number of gates and the circuit depth, which affects negatively the reliability of the computation. Therefore, minimizing this mapping overhead is crucial, especially for NISQ devices in which no or hardly any error protection mechanisms will be used. We have shown two examples of mappers developed for two specific superconducting transmon processors, the *IBM QX4* and the *Surface-17*, where different solution approaches are used. In addition, we have discussed other device types, the internal data representations used by the mappers and described the peculiarities of other possible physical implementations of quantum processors.

There are still several open questions requiring the attention of the community working on mappers of quantum algorithms. First, *what is the best metric to optimize?* Most of the works use as the optimization metric either the number of gates or the circuit depth. Recent works started considering the expected reliability of the overall quantum computation. We believe that new metrics, or possibly a combination of the existing ones, need to be investigated. Secondly, *should we aim for machine-specific solutions or more general-purpose and flexible ones capable of targeting different quantum devices and technologies and different optimization problems?* So far, the proposed mappers can be considered ad-hoc solutions that are mostly meant for a particular chip or similar kind of processors in which qubits are moved by SWAPs. While general-purpose mappers would avoid repeating the development effort for each device, the risk is that general optimization strategies will not take full advantage of the hardware capabilities. In addition, the change of the quantum technology may require very different mapping strategies. Third, *what is the good balance between the obtained solution and the time required to compile the circuit?* It is necessary to analyze the trade-off between mapping optimizations and runtime, specially for large-scale quantum algorithms. Finally, it is important to mention that these optimizations should consider both the quantum device and the quantum application characteristics. In this direction, reference [69] proposes an approach which takes the planned quantum functionality into account when determining an architecture.

## Acknowledgment

We sincerely thank all co-authors and collaborators who worked with us in the past in this exciting area. This work has partially been supported by the LIT Secure and Correct System Lab funded by the State of Upper Austria (RW) and by Intel Corporation (CGA).